\documentclass[draft,jgrga]{agutex}

\usepackage{graphicx}
\usepackage{setspace}
\usepackage{xcolor,colortbl}
\usepackage{booktabs}
\usepackage{multirow}
 \usepackage{epsfig}

\usepackage{amssymb}
\usepackage{amsmath}
\usepackage{lineno}
\usepackage[T1]{fontenc}
\usepackage[latin1]{inputenc}
\usepackage{color}
\usepackage{pgf}
\authorrunninghead{KLAUSNER ET AL.}

\titlerunninghead{TSUNAMI EFFECTS ON THE GEOMAGNETIC FIELD}

\authoraddr{Corresponding author: V. Klausner,
Physics and Astronomy Department,
Vale do Paraiba University - UNIVAP 12244-000, S\~ao Jos\'e dos Campos, SP, Brazil.
(virginia@univap.br)}

\begin{document}

%\graphicspath{{FIG_eps/}}

%% ------------------------------------------------------------------------ %%
%
%  TITLE
%
%% ------------------------------------------------------------------------ %%

\title{Rayleigh and acoustic gravity waves detection on magnetograms during the Japanese Tsunami, 2011}
%
% e.g., \title{Terrestrial ring current:
% Origin, formation, and decay $\alpha\beta\Gamma\Delta$}
%

%% ------------------------------------------------------------------------ %%
%
%  AUTHORS AND AFFILIATIONS
%
%% ------------------------------------------------------------------------ %%

\authors{V. Klausner, \altaffilmark{1,2,3}
 Esfhan A. Kherani, \altaffilmark{2}
 Marcio T. A. H. Muella, \altaffilmark{1}
 Odim Mendes, \altaffilmark{2}
Margarete O. Domingues, \altaffilmark{2}
and Andres R. R. Papa \altaffilmark{3,4}
 }
 
\altaffiltext{1}{Vale do Paraiba University, S\~ao Jos\'e dos Campos, SP, Brazil}

\altaffiltext{2}{National Institute for Space Research, S\~ao Jos\'e dos Campos, SP, Brazil.}

\altaffiltext{3}{Department of Geophysics, National Observatory, Rio de Janeiro, RJ, Brazil.}

\altaffiltext{4}{State University of Rio de Janeiro, Rio de Janeiro, RJ, Brazil.}

%Use \author{\altaffilmark{}} and \altaffiltext{}

% \altaffilmark will produce footnote;
% matching \altaffiltext will appear at bottom of page.

% \authors{A. B. Smith,\altaffilmark{1}
% Eric Brown,\altaffilmark{1,2} Rick Williams,\altaffilmark{3}
% John B. McDougall\altaffilmark{4}, and S. Visconti\altaffilmark{5}}

%\altaffiltext{1}{Department of Hydrology and Water Resources,
%University of Arizona, Tucson, Arizona, USA.}

%\altaffiltext{2}{Department of Geography, Ohio State University,
%Columbus, Ohio, USA.}

%\altaffiltext{3}{Department of Space Sciences, University of
%Michigan, Ann Arbor, Michigan, USA.}

%\altaffiltext{4}{Division of Hydrologic Sciences, Desert Research
%Institute, Reno, Nevada, USA.}

%\altaffiltext{5}{Dipartimento di Idraulica, Trasporti ed
%Infrastrutture Civili, Politecnico di Torino, Turin, Italy.}

%% ------------------------------------------------------------------------ %%
%
%  ABSTRACT
%
%% ------------------------------------------------------------------------ %%

% >> Do NOT include any \begin...\end commands within
% >> the body of the abstract.

\begin{abstract}
The continuous geomagnetic field survey holds an important potential in future prevention of tsunami damages, and also, it could be used in tsunami forecast. 
In this work, we were able to detected for the first time Rayleigh and ionospheric acoustic gravity wave propagation in the Z-component of the geomagnetic field due to
the Japanese tsunami, $2011$ prior to the tsunami arrival.
The geomagnetic measurements were obtained in the epicentral near and far-field.
Also, these waves were detected within minutes to few hours of the tsunami arrival.
For these reasons, these results are very encouraging, and confirmed that the geomagnetic field monitoring could play an important role in the tsunami warning systems, and also, it could provide
additional information in the induced ionospheric wave propagation models due to tsunamis.
\end{abstract}

%% ------------------------------------------------------------------------ %%
%
%  BEGIN ARTICLE
%
%% ------------------------------------------------------------------------ %%

% The body of the article must start with a \begin{article} command
%
% \end{article} must follow the references section, before the figures
%  and tables.

\begin{article}

%% ------------------------------------------------------------------------ %%
%
%  TEXT
%
%% ------------------------------------------------------------------------ %%

\section{Introduction}
\label{Introduction}

The monitoring of the geomagnetic field has proved to be an auxiliary tool in the study of tsunami by several authors \citep{Balasis&Mandea2007,Manoj2011,Utadaetal:2011,Klausneretal:2014}.
The response of the ionosphere to tsunamis and the Rayleigh and gravity waves induced by them has been studied broadly over the years 
\citep{Occhipinti2006,Occhipinti2008,Occhipinti2010,Occhipinti2013,Rolland2010,Rolland2011,Galvanetal:2012}, and modeled \citep{Occhipinti2011,Kheranietall2012}.

It is well-known that post-seismic acoustic-gravity and Rayleigh waves can be observable close to the epicenter (in the near-field, within $~500$ km with velocity up to $\cong 3 km/s$)
due to direct the vertical displacement of the ground induced by the rupture.
Indeed, the rupture, as a Dirac function, have a broad spectrum of energy including both, acoustic and gravity waves
as presented on the detailed simulational-observational work by \cite{Kheranietall2012} using TEC and magnetic data.
In the far-field, tsunamis induce pure gravity waves, and additionally Rayleigh waves induce 
pure acoustic waves \citep{Occhipinti2010}.
More details could be found in \cite{Occhipinti2013}.

%The ionospheric effects of the Sumatran tsunami ($2004$) was observed by \cite{DasGupta:2006} using TEC measurements at Indian stations (far-field) after $45$ minutes of the earthquake,
%and also, the TEC variations was observed and modeled by \cite{Mai&Kiang2009,Occhipinti2011,Occhipinti2013} for the same event.
The theoretical and observational evidences show that tsunamis can generate Rayleigh and acoustic gravity waves (AGWs) in the atmosphere/ionosphere by tsunami-atmosphere-ionosphere (TAI) dynamic coupling,
and this coupling effect do not modify the main frequencies of these waves.
The presence of varieties of wavefronts propagating with velocity ranging between Rayleigh to AGW velocity was observed and discussed here for the first time
using ground magnetic data for the Japanese tsunami, $2011$ in the near and far-field (distances above $~500$ km).
These observations reaffirmed the idea that the magnetogram data can be used as a possible tool for tsunami warnings as it was already discussed by \cite{Klausneretal:2014}.

\section{Dataset}
\label{Magnetic Data}

On the 11th of March at $05:46$ UT, 2011, a powerful earthquake of magnitude $8.9\;M_w$ generated a catastrophic tsunami which propagated in the Pacific Ocean.
The epicenter was centered at $38.3\,^{\circ}$N and Long. $142.4\,^{\circ}$E, near to the coast of Japan.
To study this event, we used the minutely magnetogram data from the Z-component from $9$ ground magnetic stations which are displayed on Table~\ref{table:ABBcode}.

\section{Methodology}
\label{Methodogy}

In the work of \cite{Klausneretal:2014}, the wavelet analysis was proofed to be an alternative tool in detection of magnetic fields induced by the tsunami propagation.
Nowadays, the discrete wavelet transform (DWT) has been used in many different works in geophysics \citep{Domingues2005, MendesMag2005, MendesdaCostaetal:2011, Klausneretal:2014,Klausneretal:2014b}.
The DWT is based on the multi-scale analysis and local regularities of the signal.
Considering $\psi$ as the analyzing wavelet, the wavelet coefficients of these transforms are dependent of two parameters: the scale $a$ 
and the central position of the wavelet analyzing function translation $b$.

Here, the orthogonal discrete wavelet transform was used \citep{Daubechies1992}.
The key aspect of this transform is that the amplitude of the wavelet coefficients can be associated to the local polynomial approximation error which
can be defined by the choose of the analyzing wavelet.

Using the squared modulus of the wavelet coefficients $|d^j_k|^2$, we can reproduce the discrete scalogram related to a discrete scale and a position,
where the scale is related to a dyadic decomposition in levels $j$, as $a=2^j$ and the translation is related to discrete position  $b=2^{-j} k$  at level $j$ with $k \in \mathbb{Z}$.

In this context, we have that a signal $f(t)$ can be represented by the expansion
\begin{equation}
 f(t)=\sum_{j=-\infty}^{\infty}\sum_{k=-\infty}^{\infty}d_k^j\psi_k^j(2^jt-k),
\end{equation}
\noindent
where $d_k^j$ are the wavelet coefficients computed from the $L^2$ inner-product
\begin{equation}
 d_k^j=\int f(t)\,\psi_k^j(2^jt-k)\,dt.
\end{equation}

The wavelet transform in level $j+1$ is given by
  \begin{align} 
              d_k^{j+1} &= 2 \sum\limits_m g(m-2k) \; c_m^{j},
   \end{align}
\noindent
where  $g$ is a high-pass filter, $d_k^{j+1}$ is the wavelet coefficient at level $j+1$, $c_m^{j}$ are the scale coefficients at level $j$, and $m \in \mathbb{Z}$ \citep[see][more details]{Domingues2005}.

In this work, we use the Daubechies (db2) wavelet function of order 2, and 
therefore the non-zero low filter values are $h\cong [\frac{1+\sqrt{3}}{4\sqrt{2}}, \frac{3+\sqrt{3}}{4\sqrt{2}}, \frac{3-\sqrt{3}}{4\sqrt{2}},\frac{1-\sqrt{3}}{4\sqrt{2}}]$.
Also, the sampling rate of $1$ min, as consequence of the time resolution of the magnetic data, 
gives the pseudo-periods of the first three levels of $3, 6$ and $12$ minutes, \citep[see][more details]{Klausneretal:2014b}.

\section{Results and analysis}
\label{Results and analyses}

\begin{figure}[htb]
\begin{tabular}{c c}
 \centering
		\includegraphics[height=7cm,width=7cm]{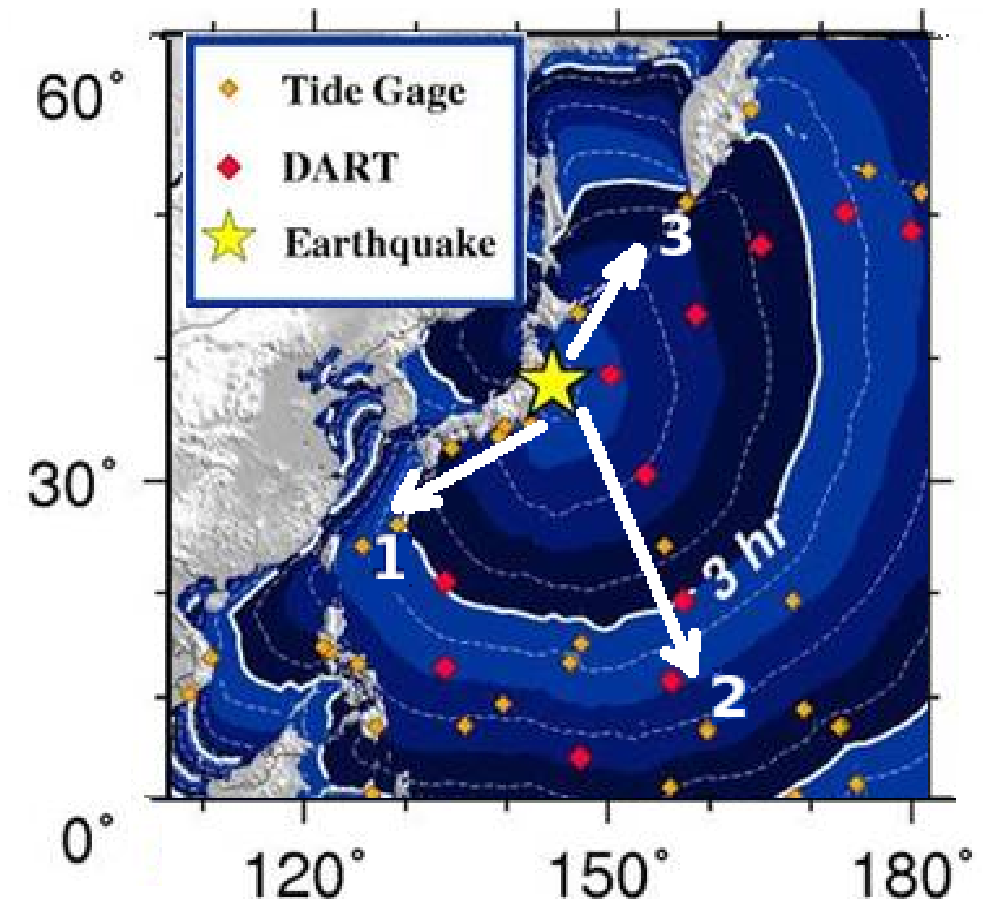}&
		\includegraphics[height=7cm,width=7cm]{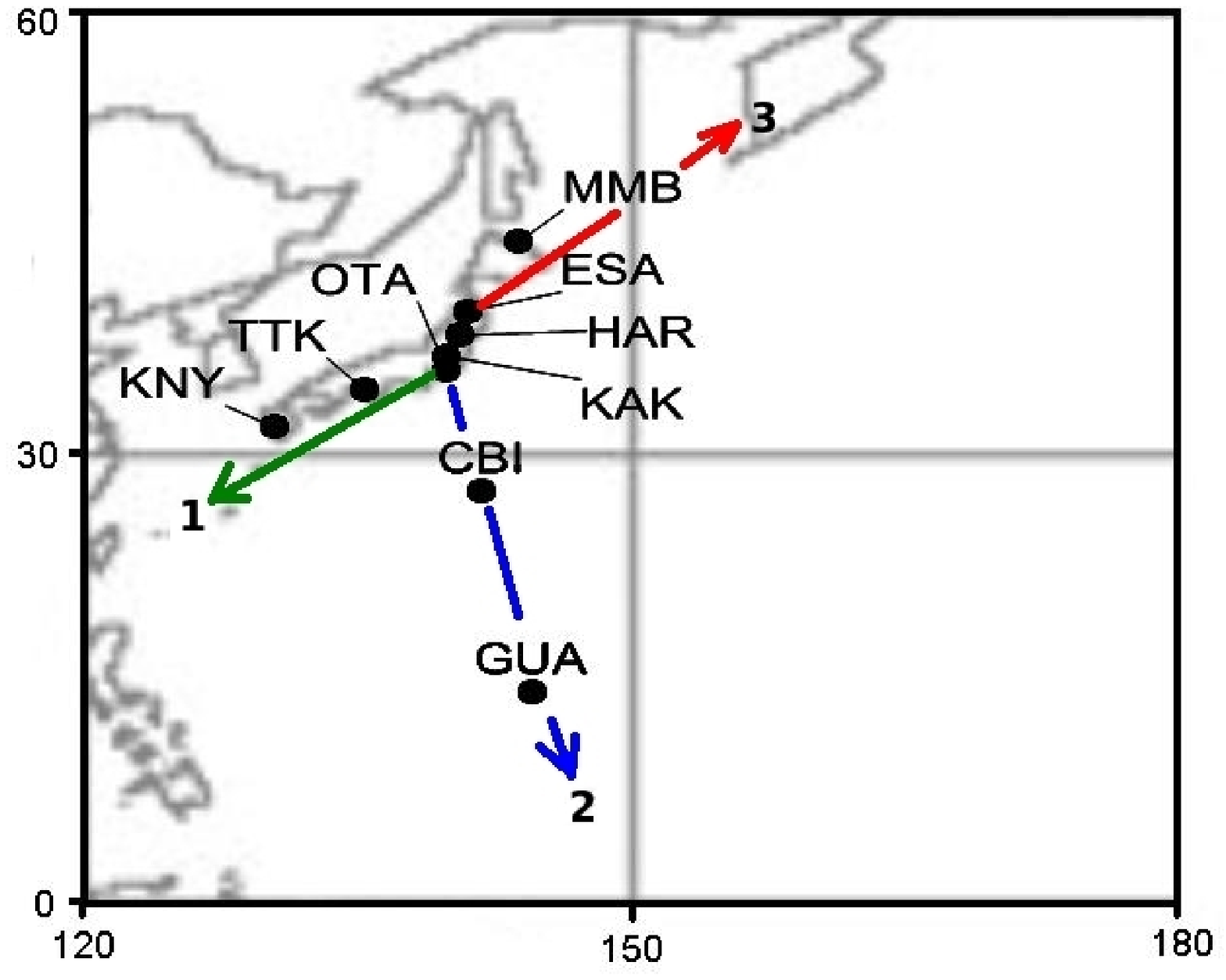}\\
\end{tabular}
	\label{fig:MapJapStations}
	\caption{TTT map and map of the geographic localization of the magnetic observatories. 
	TTT map courtesy of NOAA/NWS/West Coast and Alaska Tsunami Warning Center.}
\end{figure}

Figure~\ref{fig:MapJapStations} displays, on the right side panel, the magnetic observatories geographic distribution, and  on the left side panel, the Tsunami Travel Times (TTT) map.
In both graphics, the arrows displays the observatories used in three different wave front of the tsunami propagation direction that we use here to study the tsunamigenic disturbances,
and these directions are denoted by the numbers $1$, $2$ and $3$. 
Also for guiding purposes, we suggest the use of TTT map to determine the approximated tsunami time arrival for each magnetic observatories.

\begin{figure}[ht]
	\centering
		\includegraphics[width=16cm]{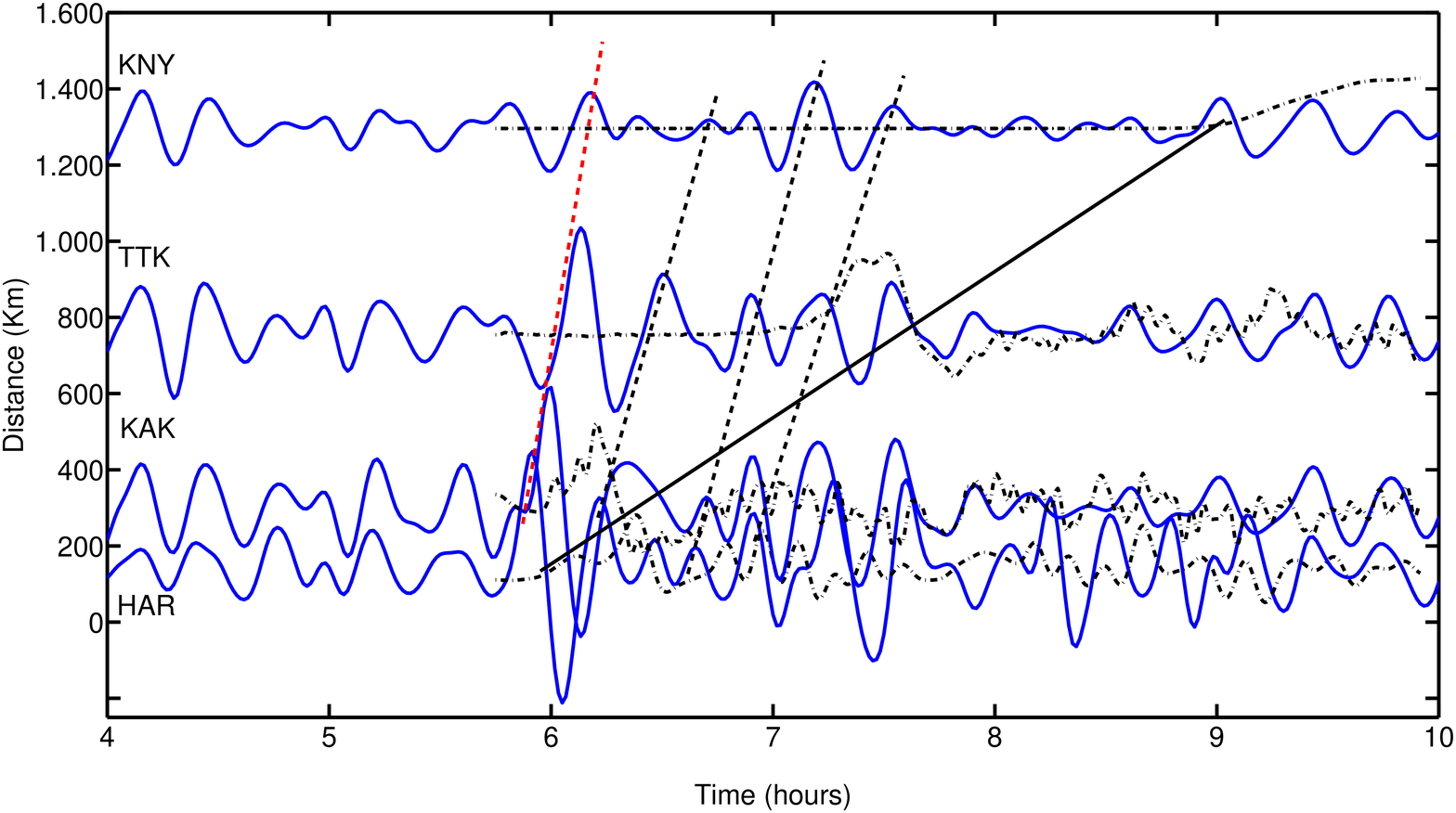}\\
		\label{fig:TTD_Thai}
	\caption{Z-component dataset filtered with a bandpass with periods between 10 and 30 minutes. On the vertical axis, it is shown the distance between the earthquake epicenter and the magnetic observatory,
	and on the horizontal axis, the universal time. The continuous line represents the tsunami waves propagation at  $\cong 200 m/s$, 
	the red dashed line is associated to Rayleigh waves with velocity of $\cong 1.3 km/s$, and
	the black dashed lines are associated to AGWs with velocity of $\cong 640 m/s$. }
\end{figure}

In the sequence, we will present the analysis of each wave front direction case study.
The wave front direction number $1$ is parallel to the Japanese coast in the west-southwest direction, and in this case the magnetic observatories used are HAR, KAK, TTK and KNY.
Figure~\ref{fig:TTD_Thai} shows the travel-time diagram (TTD) of a magnetic induced field trends due to the tsunami in the direction $1$. 
As done previous by \cite{Kheranietall2012}, the magnetograms of HAR, KAK, TTK and KNY (blue color) have been filtered using a Butterworth filter with 10 to 30 minutes period bandwidth 
to extract the tsunami-related magnetic induced fields from the external ionospheric activity.
Also, the tsunami wave propagation was simulated using the simulation model developed by \cite{Sladenetal:2007,Sladen&Herbert:2008} (black dot-dashed line).
The continuous line represents the tsunami wave front propagation with velocity of $\cong 200 m/s$, the red dashed line is associated to Rayleigh waves propagation ($\cong 1.3 km/s$),
and the dashed lines are associated to acoustic gravity waves propagation $\cong 640 m/s$.
It is well-known that tsunami waves can induce long-wavelength acoustic gravity waves (AGWs) with velocity up to $\cong 800 m/s$ 
and short-wavelength gravity waves with velocity of $\cong 200 m/s$ \citep{Kheranietall2012},
and additionally Rayleigh waves with velocity up to $\cong 2.1 km/s$ which can induce pure acoustic waves \citep{Occhipinti2010}.
In Figure~\ref{fig:TTD_Thai}, Rayleigh and acoustic gravity waves fronts is observed in the presented TTD.
It is possible to notice that depending in the distance from the epicenter the acoustic gravity waves can be detected within minutes to hours before the arrival of the tsunami wave.
By this reason, we can discriminate the effect (on the magnetic data) induced by the ionospheric perturbation, 
from the effect induced by the secondary magnetic field induced directly by the displacement of the oceanic surface (the tsunami). 

\begin{figure}[H]
\noindent
\centering 
\begin{tabular}{c}
\includegraphics[width=9cm]{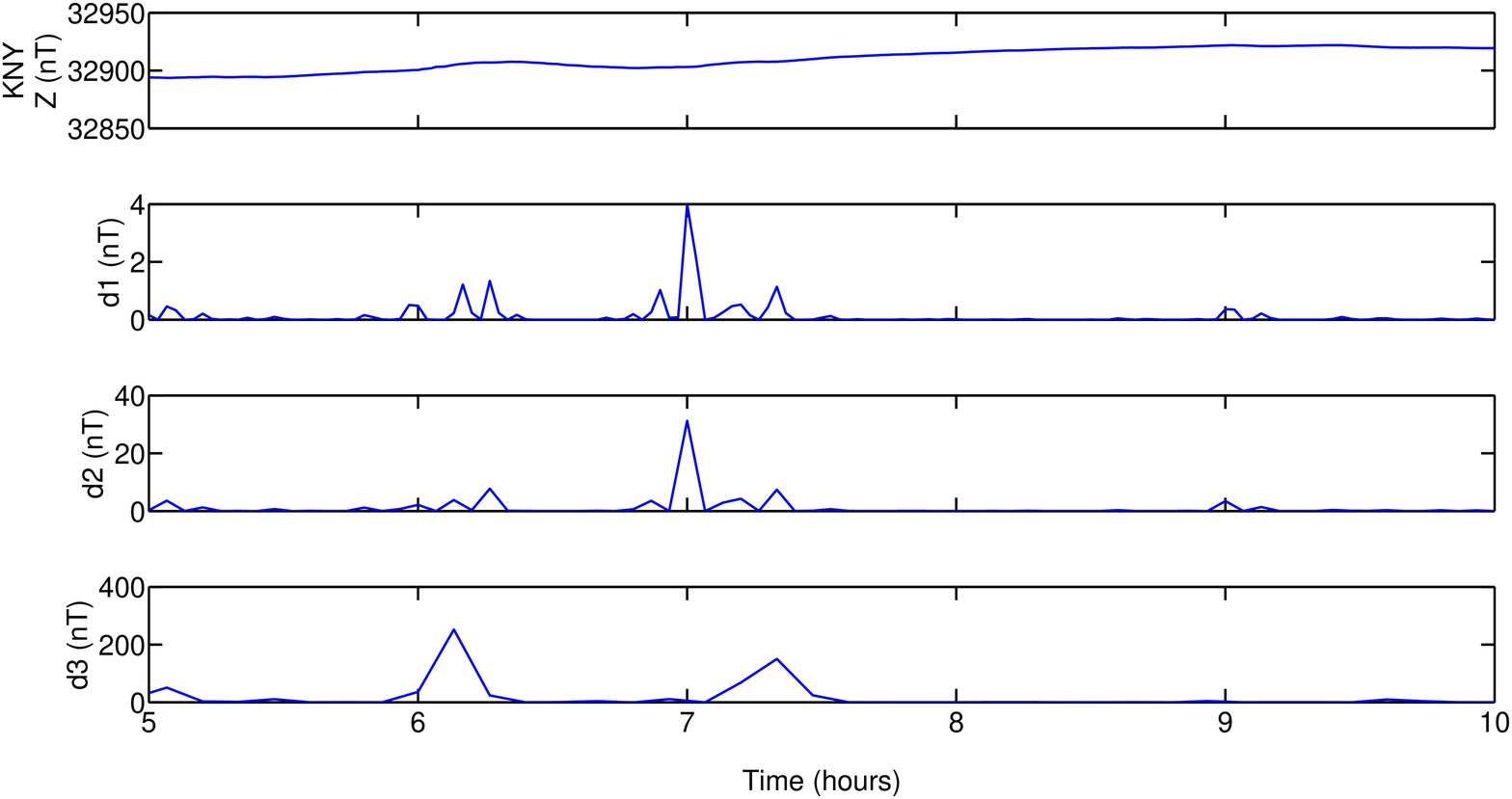}\\
\includegraphics[width=9cm]{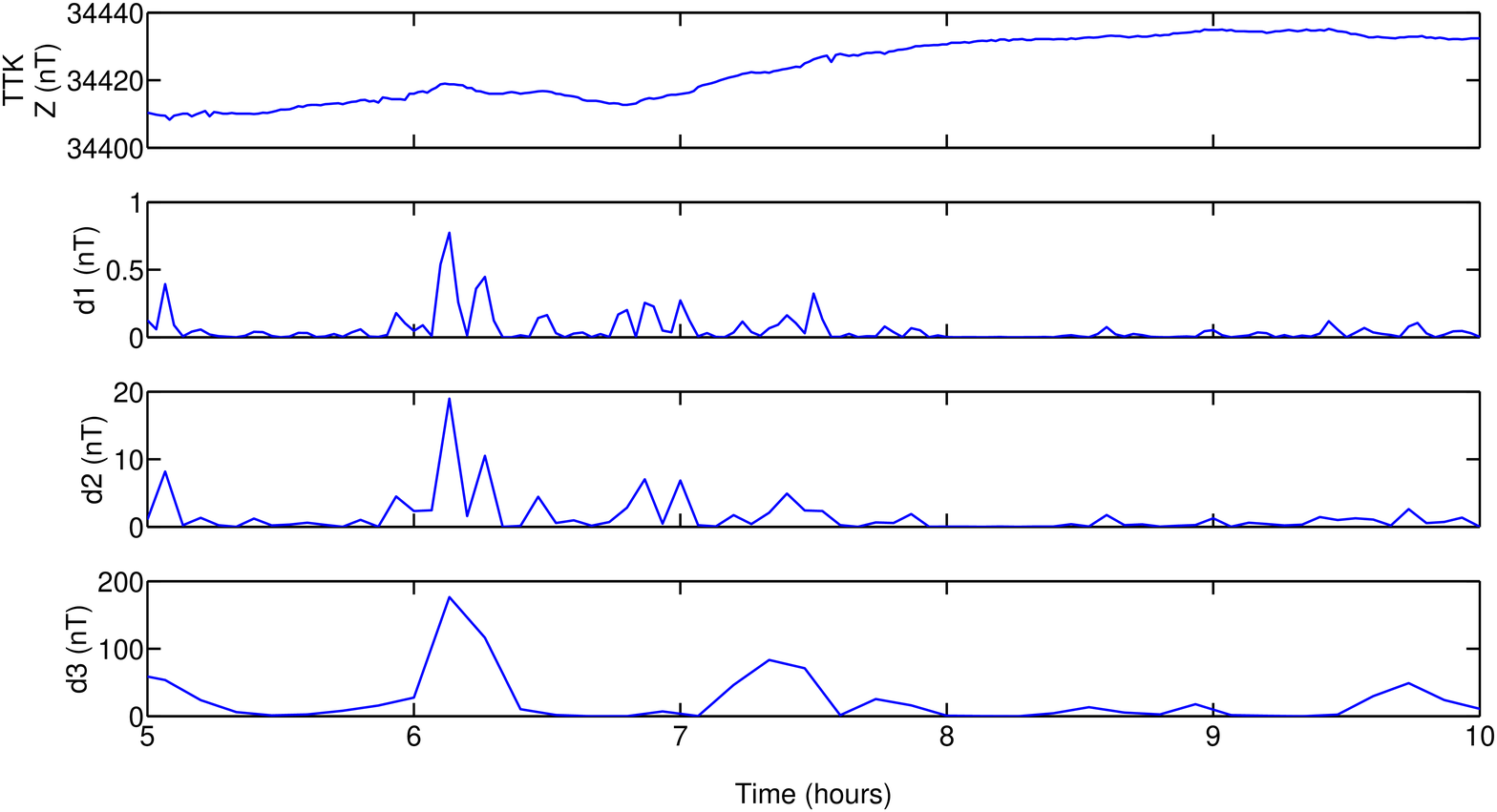}\\
\includegraphics[width=9cm]{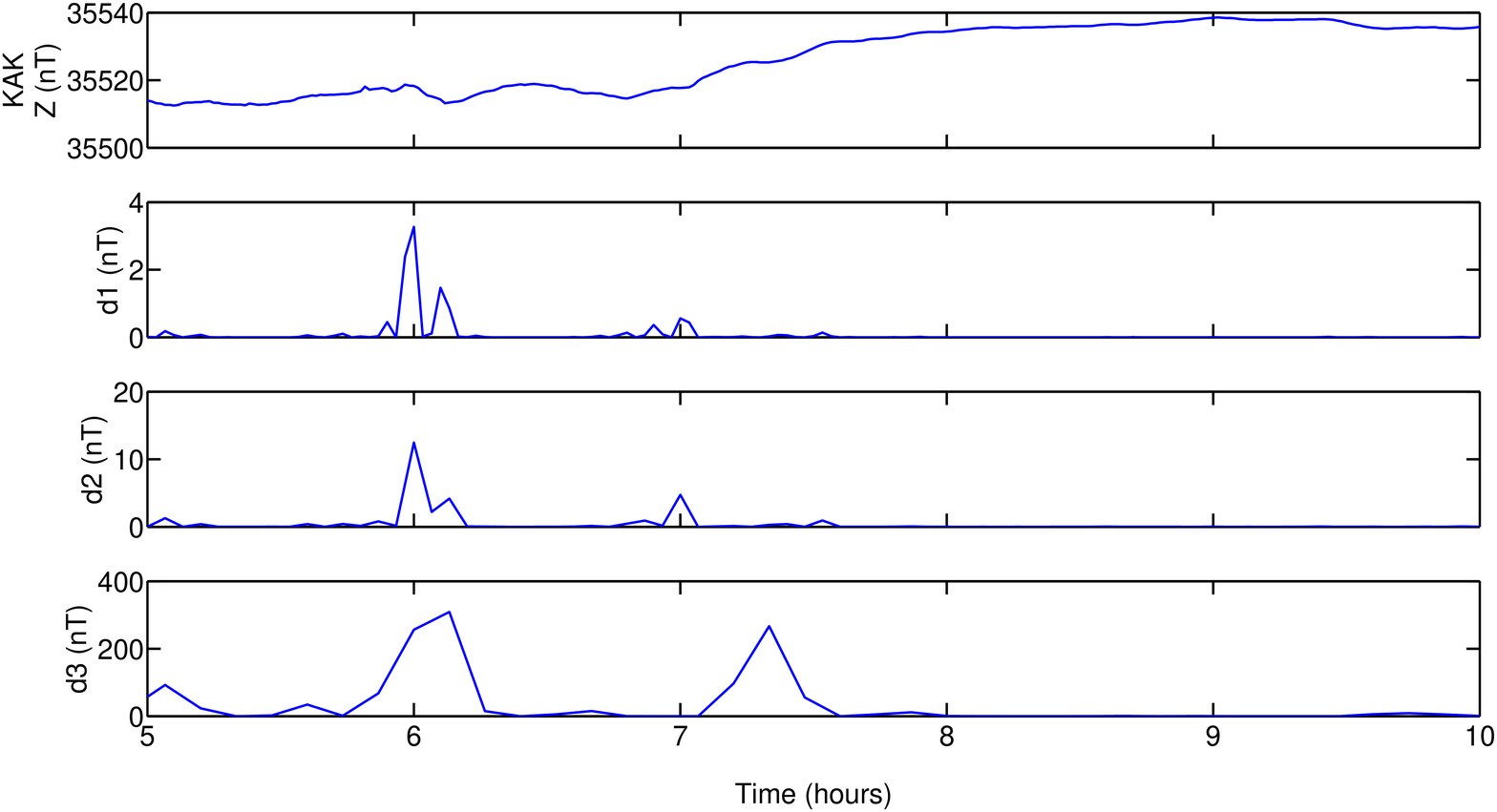}\\
\includegraphics[width=9cm]{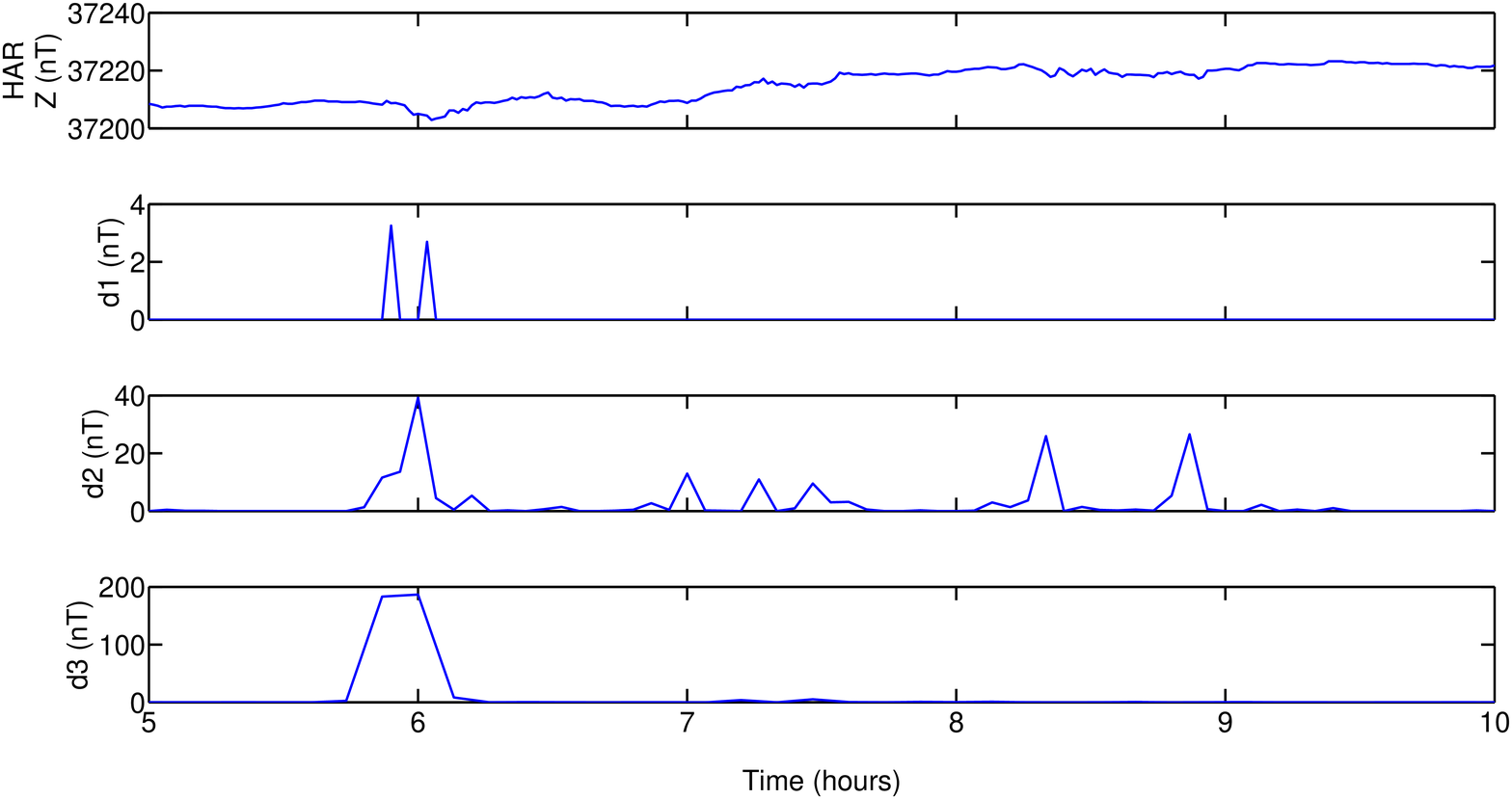}\\
\end{tabular}
\caption{Magnetograms and the wavelet coefficients $(d^j)^2$ for $j=1,2,3$ for wave front case study in the direction $1$.}
\label{fig:11MarDecomposition}
\end{figure}

In Figure~\ref{fig:11MarDecomposition}, panels from bottom (near-field) to top (far-field) correspond to magnetic observatories of HAR, KAK, TTK and KNY, respectively.
Each panel, from top to bottom, displays the corresponding magnetogram (Z component), and
the square wavelet coefficients for the three first decomposition levels, $(d^j)^2$ for $j=1,2,3$.
It is possible to notice that Figure~\ref{fig:11MarDecomposition}
shows  an increase of wavelet coefficient amplitudes (WCAs) before of the tsunami arrival which is owing to the contribution from both geomagnetic disturbances and tsunami
induced ionospheric disturbances.
In particular, following features are evident from Figure~\ref{fig:11MarDecomposition}: 
(a) The wavelet coefficients are amplified after few minutes of the earthquake, and
(b) At TTK and KNY, amplifications are noted few minutes before the tsunami arrival.
These features suggest that the observed disturbances in a few minutes after the earthquake may be associated to Rayleigh,
and the amplifications  few minutes before the tsunami arrival may be associated to the acoustic ionospheric gravity propagation, respectively. 
However, we can only come to this conclusion if these identified tsunamigenic disturbances are observed also in the TTD.
This kind of analysis makes possible to say that these disturbances can be classified to be from seismic and from tsunami origin.

In a near epicentral field (HAR and KAK), the delay between the tsunami and the AGWs is always positive, see \cite{Occhipinti2013} for more details.
And in the far-field (TTK and KNY), the AGWs follows the Rayleigh waves arrival, and they can be detected even with $2$ hours in advance of the tsunami arrival.

\begin{figure}[ht]
	\centering
		\includegraphics[width=16cm]{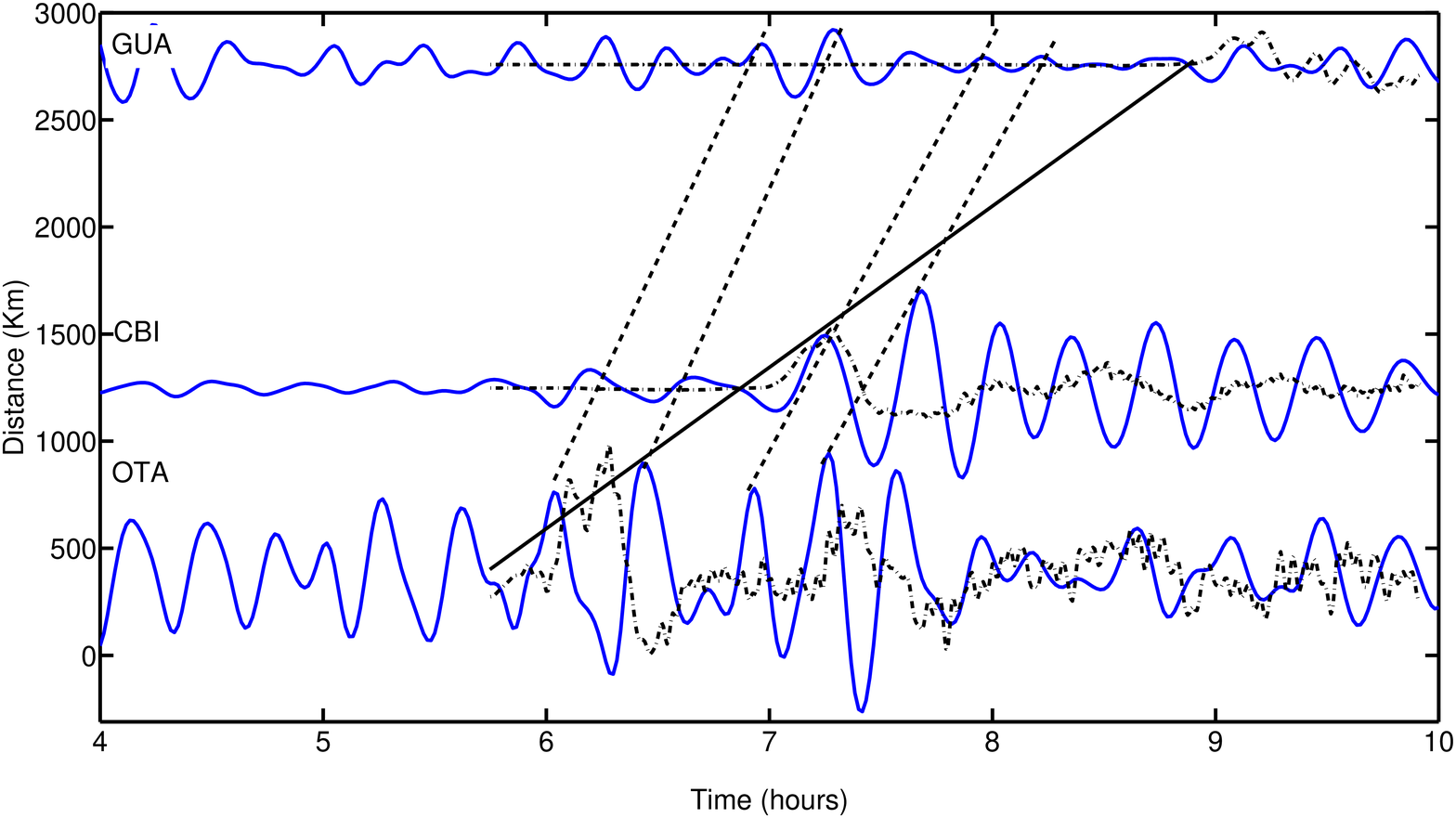}\\
		\label{fig:TTD_2}
	\caption{Same as Fig.~\ref{fig:TTD_Thai} but for the direction $2$.}
\end{figure}

The TTD for the direction $2$ is shown in Figure~\ref{fig:TTD_2} in which the AGWs
disturbances are identified following the same strategy as described in the context of Figure~\ref{fig:TTD_Thai}.
The wave front number $2$ is towards the open sea waters, and in this case the magnetic observatories used are OTA, CBI and GUA.
It can be again said that such identification classifies these disturbances
to be associated to AGWs wave propagation ($\cong 700 m/s$) in which appear within a few minutes to hours before the tsunami arrival.

 \begin{figure}[H]
\noindent
\centering 
\begin{tabular}{c}
\includegraphics[width=9cm]{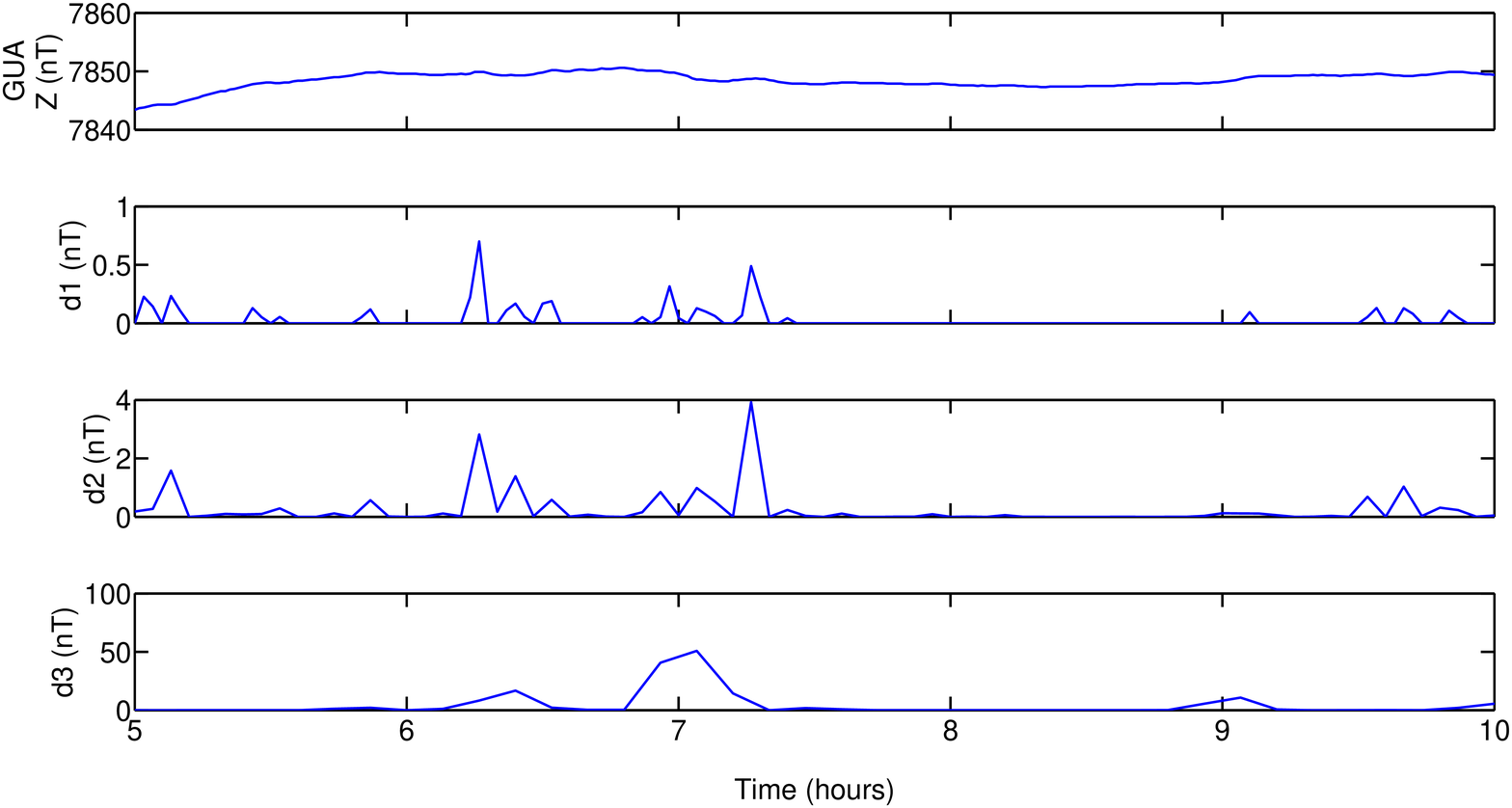}\\
\includegraphics[width=9cm]{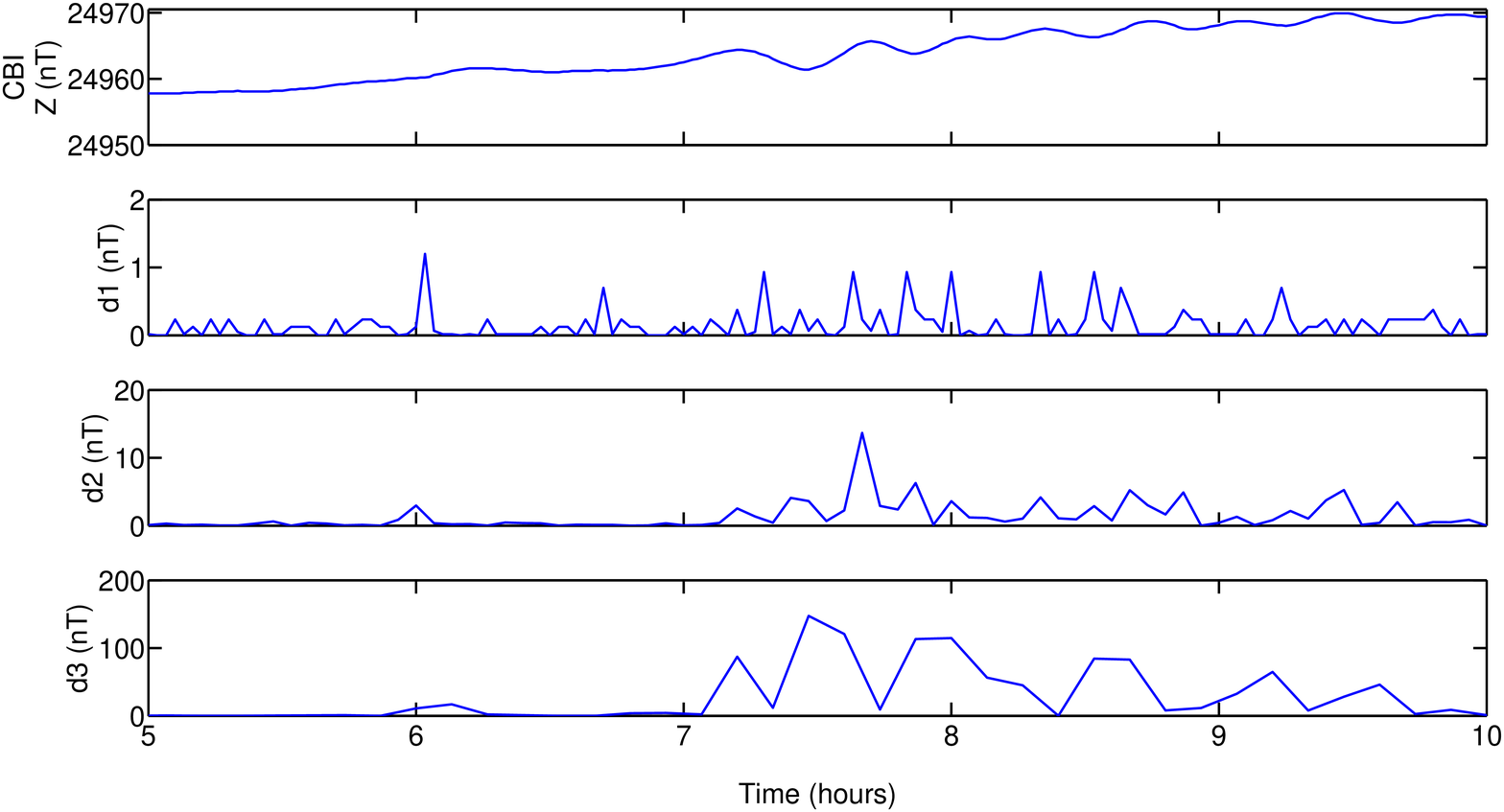}\\
\includegraphics[width=9cm]{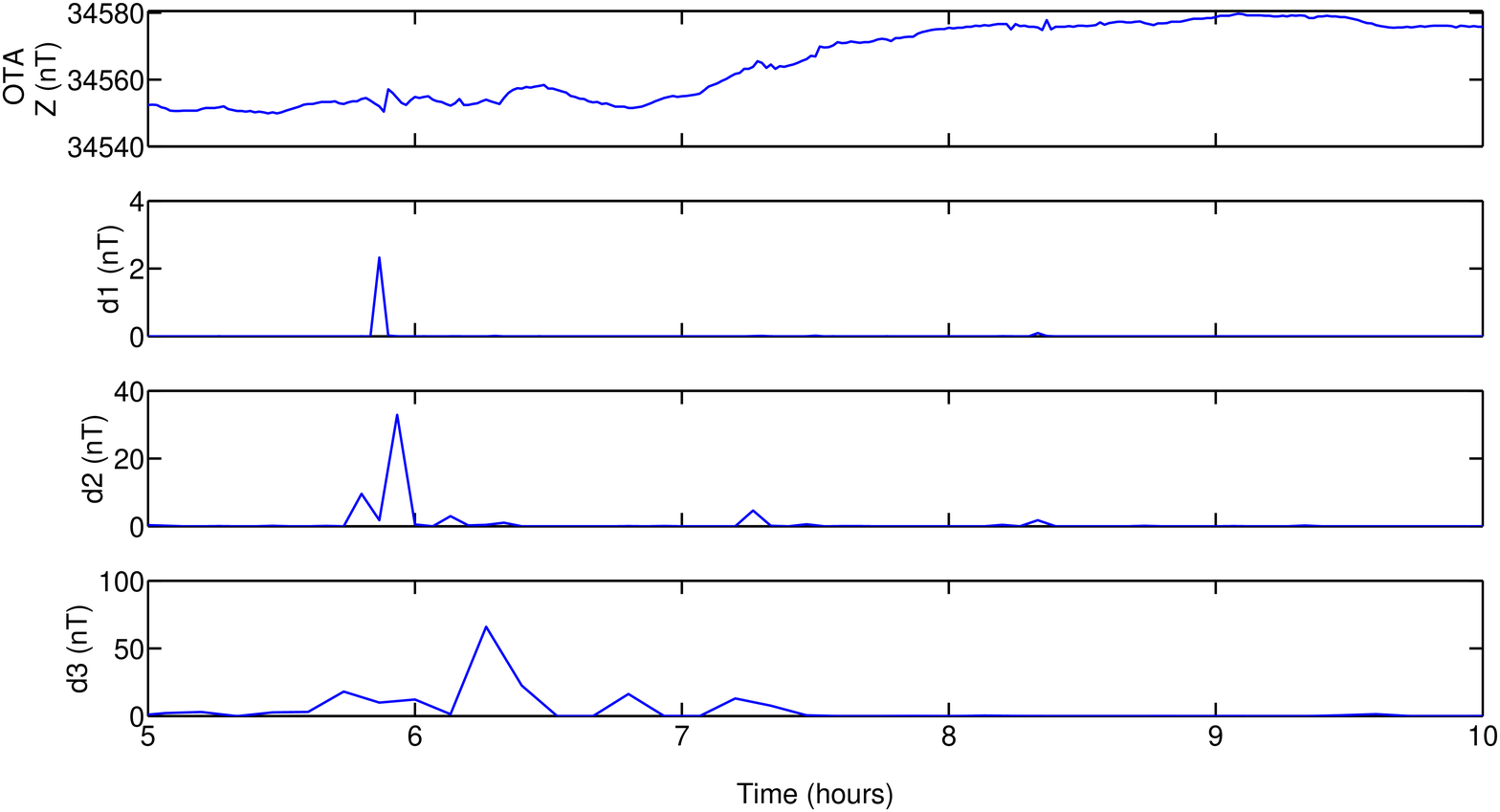}\\
\end{tabular}
\caption{Same as Fig.~\ref{fig:11MarDecomposition} but for the direction $2$.}
\label{fig:DWT2}
\end{figure}

In Figure~\ref{fig:DWT2}, the vertical component and corresponding three discrete wavelet coefficients for the magnetic observatories of OTA, CBI and GUA as indicated in Figure~\ref{fig:MapJapStations}
for the direction $2$.
We note from Figure~\ref{fig:DWT2} that the magnetic disturbances within a few minutes of the earthquake are amplified at all observatories,
these amplifications are only due to AGW propagation.
In OTA and CBI, the WCAs around $6:00$ UT may or not be associated to the Rayleigh wave propagation.
However, these features related to the Rayleigh wave propagation were not detected in the TTD (Figure~\ref{fig:TTD_2}).
Also at CBI, the tsunamigenic disturbances after the tsunami arrival are associated to AGW propagation.
And these waves have propagation characteristics similar to the tsunami wavefront propagation.
For example, the predicted time of the tsunami's arrival by NOAA at Chichijima Island was at $07:14$ UT. 
The tide gauge measurements registered the tsunami maximum height at $07:46$ UT with amplitudes up to $1.80~\mathrm{m}$.
The CBI magnetic observatory shows remarkable wavelet coefficient amplitudes $(d^j)^2$($j=1,2,3$) due to the tsunami after the tsunami arrival time as well.
These increase in WCAs correspond to the AGWs propagation.
Also, \cite{Utadaetal:2011} discussed the similarity between the CBI Z-component fluctuations and the IPM fluctuations reported by \cite{Manoj2011} due to the AGW point of view.

\begin{figure}[ht]
	\centering
		\includegraphics[width=16cm]{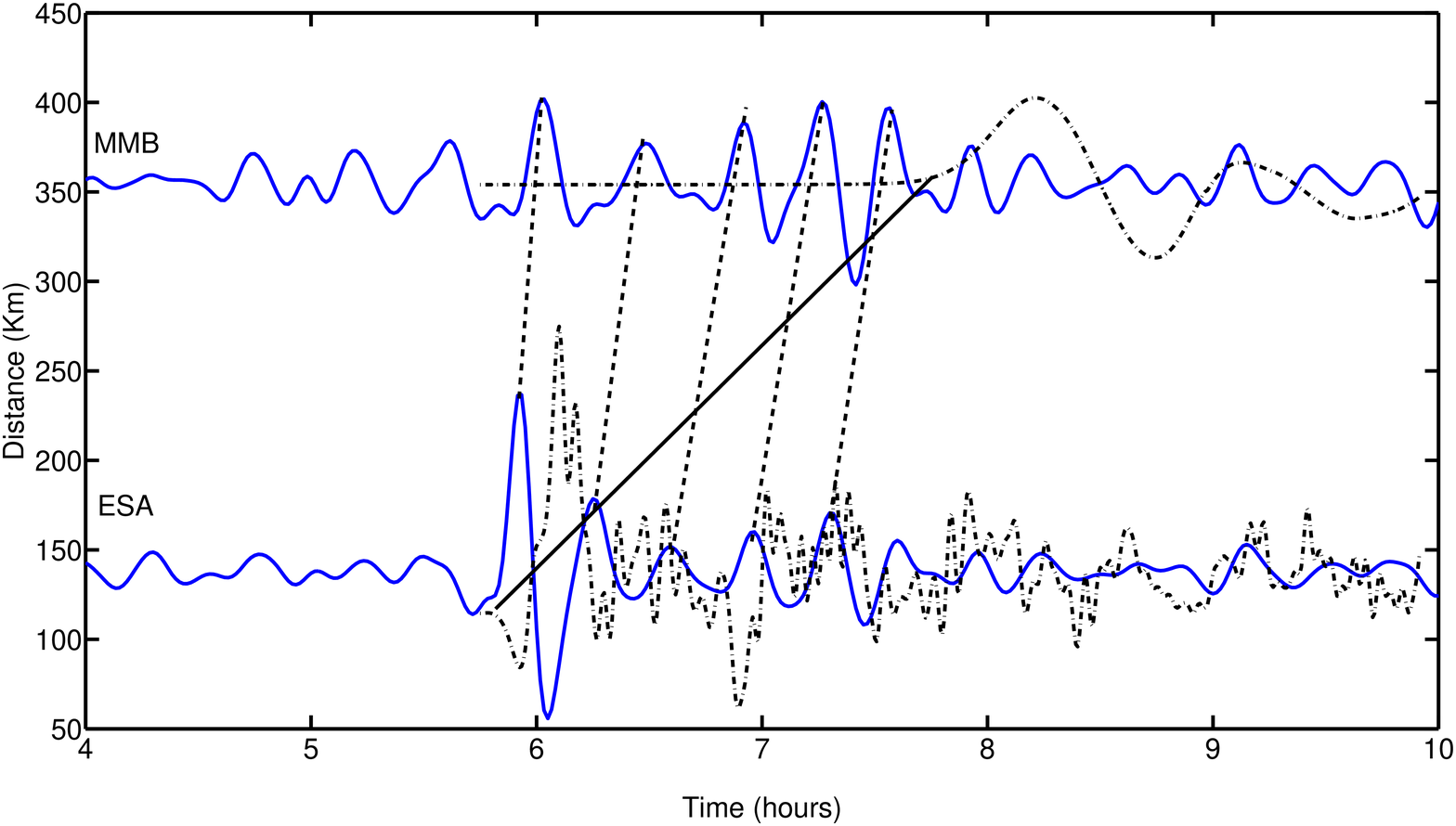}\\
		\label{fig:TTD_3}
	\caption{Same as Fig.~\ref{fig:TTD_Thai} but for the direction $3$.}
\end{figure}

\begin{figure}[H]
\noindent
\centering 
\begin{tabular}{c}
\includegraphics[width=9cm]{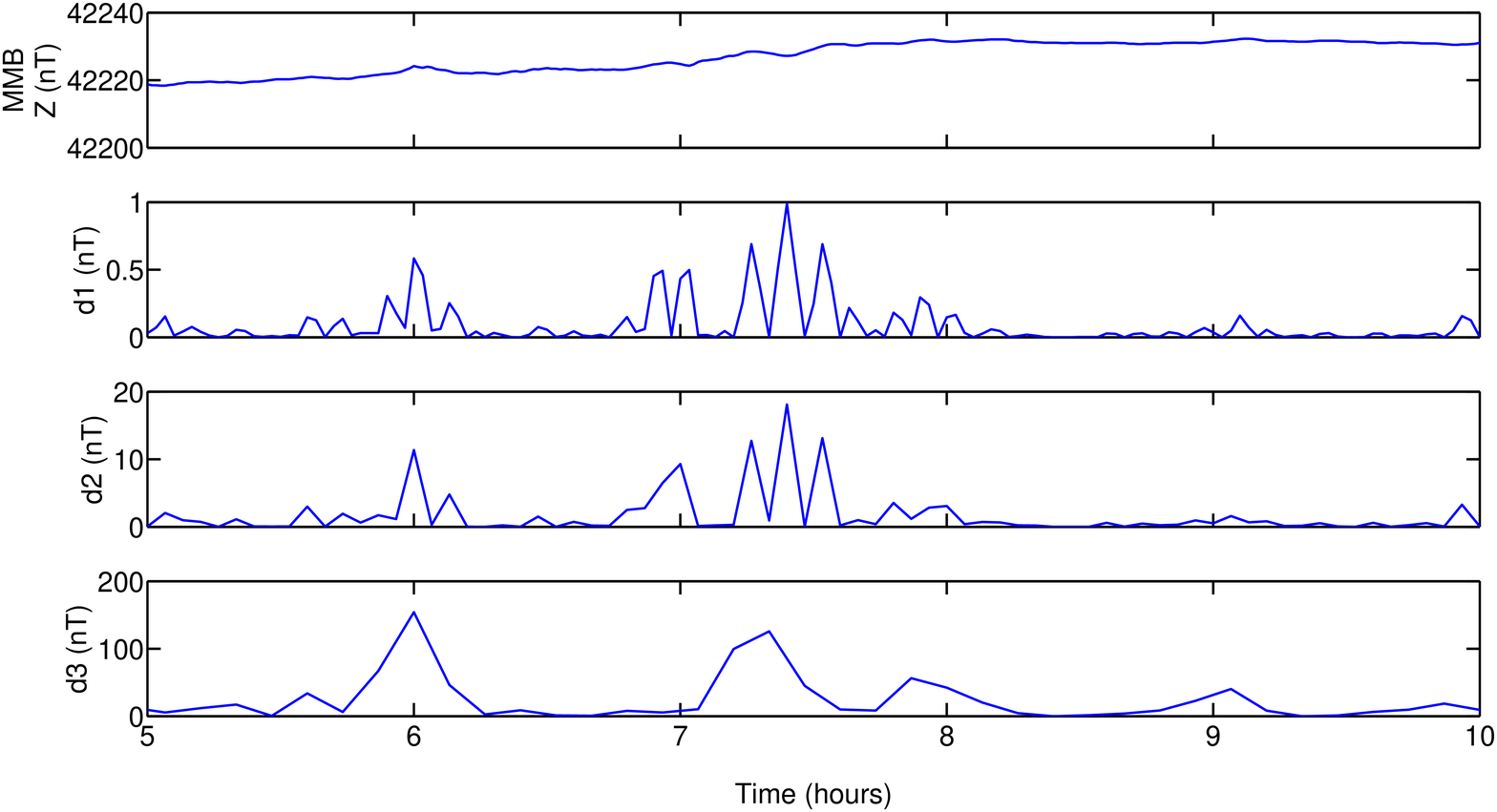}\\
\includegraphics[width=9cm]{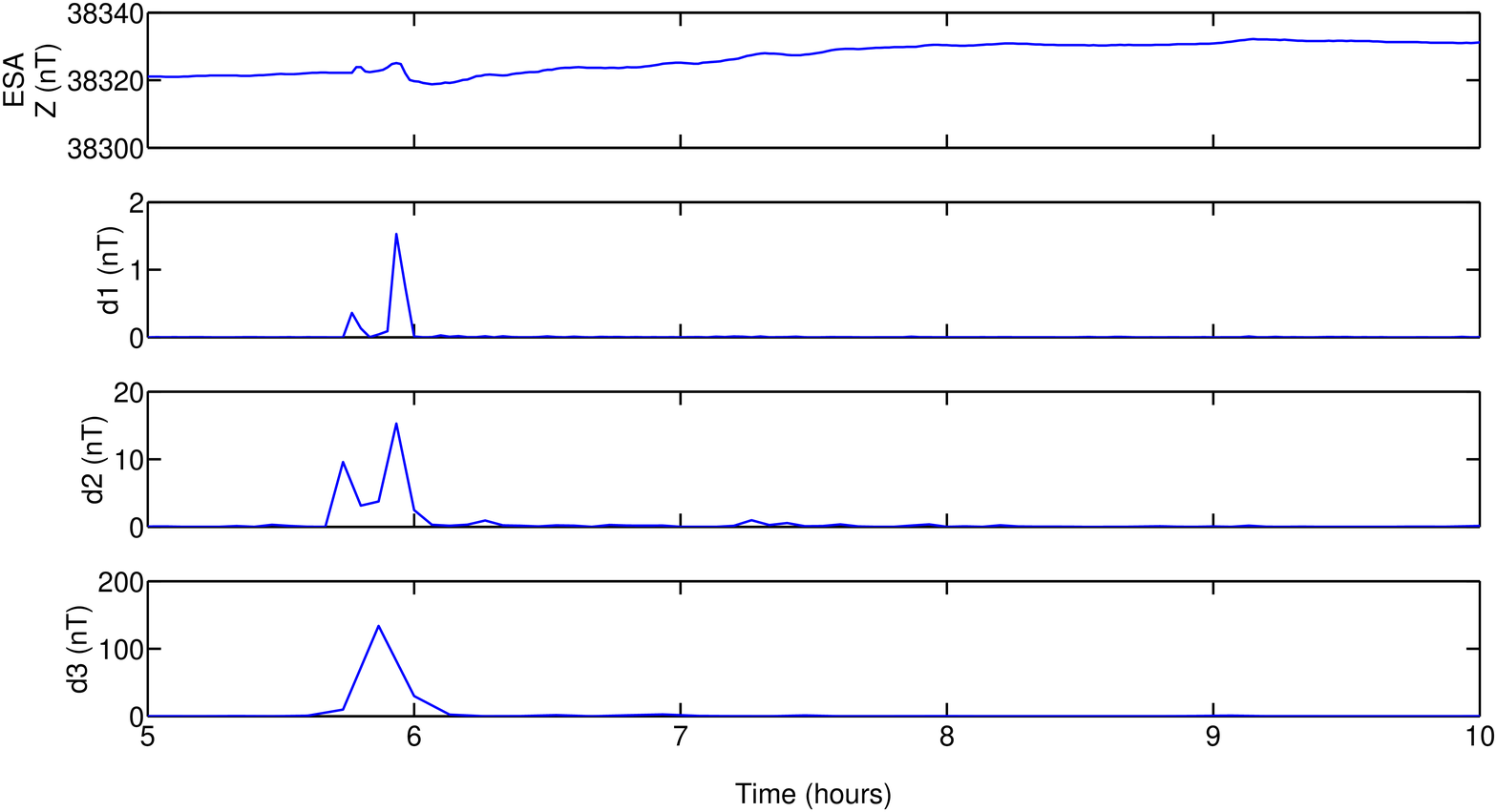}\\
\end{tabular}
\caption{Same as Fig.~\ref{fig:11MarDecomposition} but for the direction $3$.}
\label{fig:DWT3}
\end{figure}

The TTD for the direction $3$ is shown in Figure~\ref{fig:TTD_3} in which the
AGWs disturbances ($\cong 800 m/s$) are identified following again the same
strategy as mentioned in the context of Figure~\ref{fig:TTD_Thai}.
In this case, the wave front number $3$ is parallel to the Japanese coast, and the magnetic observatories used are ESA and MMB.
From Figure~\ref{fig:DWT3}, it can be again said that the magnetic disturbances due to AGWs within a few minutes of the earthquake are amplified at all observatories,
while the amplification due to AGWs could be see at MMB in $2$ hours in advance of the tsunami arrival.
The reason for that is the ESA observatory is near to the epicenter, and an increase of WCA is almost simultaneous to tsunami wave arrival.
On the other hand, near to the magnetic observatory of MMB, the initial phase of the tsunami started at $06:38$ UT and reached the maximum height ($0.74$ m) at $06:57$ UT.
The $(d^j)^2$($j=1,2,3$) showed WCAs around $06:00$ UT which may be due to Rayleigh wave propagation, and
again between $07:00$ and $08:00$ UT which might be due to AGW propagation.
At $06:57$ UT, when tsunami reached the maximum height, the presence of an increase in the coefficient amplitudes around this time might be due to tsunami maximum height too.

In all the analysis presented above, post-seismic acoustic-gravity and Rayleigh waves were observable close to the epicenter (near-field, within $~500$ km with velocity up to $\cong 2 km/s$)
due to direct the vertical displacement of the ground induced by the rupture.
Indeed, the rupture, as a Dirac function, have a broad spectrum of energy including both, acoustic and gravity waves. 
Here, the presence of varieties of wavefronts propagating with velocity ranging between Rayleigh to gravity wave velocity was observed 
as presented on the detailed simulational-observational work by \cite{Kheranietall2012} using TEC and magnetic data.
However, the Rayleigh wave propagation was only detected here in direction $1$, while AGWs were detected in all directions.
This assymetry in the distribution of Rayleigh wave propagation was also observed by \cite{Galvanetal:2012,Kakinamietal:2013} in the TEC data.
Their studies also found the Rayleigh waves propagating only in the southward direction,
although these waves should propagate in all directions.

\section{Conclusions}
\label{Conclusions}

Using filtered magnetograms of the geomagnetic Z-component, and also, the spectral-time diagrams,
we presented a detailed TTD with a presence of Rayleigh and AGWs which was not done before for the Japanese tsunami using magnetogram data, $2011$.
These perturbations presented velocities of $1.3$ km/s and $600$ -- $800$ m/s, respectively, with also it is consistent which tsunami-atmosphere-ionosphere coupling theory.
These results are in confirmation with the previous studies \citep{Occhipinti2006,Occhipinti2008,Occhipinti2010,Occhipinti2013,Rolland2010,Rolland2011,Galvanetal:2012},
however they show new insight into the tsunamigenic magnetic disturbances due to Rayleigh and AGWs on near and far-field distances from the epicenter.

As explained by \cite{Kheranietall2012}, the presence of periodicity of order of $8$ to $30$ minutes in the spectral-time diagrams and, Rayleigh and acoustic gravity waves fronts in the TTDs
are the aspects of this work which distinguish the magnetic disturbances of tsunami in the ionosphere from other external magnetic perturbation sources.
Also, \cite{Kheranietall2012} have presented the TTD and reported the presence of acoustic disturbances that remained confined to $<5^o$ which is ~$500$km distance from the epicenter.
However, they did not see any acoustic disturbance beyond this distance.
In this work, we presented these disturbances beyond 500 km from the epicenter distance and up to $3$ hours of the tsunami arrival.

Also, the fact of Rayleigh waves and AGWs can be detected with a few hours in advance of the tsunami is very encouraging, and
shows that the constant monitoring of the geomagnetic field could play an important role to tsunami forecast and tsunami early warning systems.

%%% End of body of article:

%%%%%%%%%%%%%%%%%%%%%%%%%%%%%%%%
%% Optional Appendix goes here
%
% \appendix resets counters and redefines section heads
% but doesn't print anything.
% After typing  \appendix
%
% \section{Here Is Appendix Title}
% will show
% Appendix A: Here Is Appendix Title

%
%%%%%%%%%%%%%%%%%%%%%%%%%%%%%%%%%%%%%%%%%%%%%%%%%%%%%%%%%%%%%%%%
%
% Optional Glossary or Notation section, goes here
%
%%%%%%%%%%%%%%
% Glossary is only allowed in Reviews of Geophysics
% \section*{Glossary}
% \paragraph{Term}
% Term Definition here
%
%%%%%%%%%%%%%%
% Notation -- End each entry with a period.
% \begin{notation}
% Term & definition.\\
% Second term & second definition.\\
% \end{notation}
%%%%%%%%%%%%%%%%%%%%%%%%%%%%%%%%%%%%%%%%%%%%%%%%%%%%%%%%%%%%%%%%
%
%  ACKNOWLEDGMENTS

\begin{acknowledgments}
V. Klausner wishes to thank CAPES for the financial support for her Postdoctoral research within the Programa Nacional de P\'os-Doutorado (PNPD -- CAPES)
and (FAPESP -- grants 2011/21903-3, 2011/20588-7 and 2013/06029-0).
The authors would like to thank the NOAA, GIS and the INTERMAGNET program for the datasets used in this work.
\end{acknowledgments}

\end{article}

\newpage

%% Enter Fig.ures and Tables here:

% When submitting articles through the GEMS system:
% COMMENT OUT ANY COMMANDS THAT INCLUDE GRAPHICS.
%
% DO NOT USE \psfrag or \subfigure commands.
%
% Fig.ure captions go below the figure.
% Table titles go above tables; all other caption information
%  should be placed in footnotes below the table.

% DRAFT figure/table, including eps graphics
%
\small

\begin{table}[ht]
 \caption{Network of geomagnetic observatories used in this work.}
\centering
\begin{tabular}{c c c c c }
\hline
IAGA code&Station & country& \multicolumn{2}{c}{Geographic coord.} \\
\cline{4-5}
          &  &  & Lat.($\,^{\circ}$) & Long.($\,^{\circ}$)  \\[0.5ex]
\hline
\\
CBI$^1$	& Chichijima			   &  Japan					&27.10  	& 142.19 \\
ESA$^1$ 	&Esashi 			&Japan					& 39.24			& 141.36\\     
GUA 	&Guam 	&United States of America    &13.59     &144.87  \\
HAR$^1$     &Haramachi       &Japan & 37.62      &140.95\\   
KAK 	&Kakioka 	&Japan    &36.23    &140.18      \\
KNY 	&Kanoya 	&Japan     &31.42     &130.88        \\
MMB 	&Memambetsu &	Japan     &35.44     &144.19        \\ 
OTA$^1$     & Otaki  &Japan & 35.29  &140.23\\               
TTK$^1$     &Totsugawa  & Japan  &33.93 &135.80 \\[1ex]
\hline
\end{tabular}
\\
\footnote{a}{Observatories belonging to Geospatial Information Authority of Japan (GIS).}
\label{table:ABBcode}
\end{table}

%

%
% \begin{table}
% \caption{}
% \end{table}
%
% ---------------
% TWO-COLUMN figure/table
%
% \begin{figure*}
% \noindent\includegraphics[width=39pc]{samplefigure.eps}
% \caption{Caption text here}
% \end{figure*}
%
% \begin{table*}
% \caption{Caption text here}
% \end{table*}
%
% ---------------
% EXAMPLE TABLE
%
%\begin{table}
%\caption{Time of the Transition Between Phase 1 and Phase 2\tablenotemark{a}}
%\centering
%\begin{tabular}{l c}
%\hline
% Run  & Time (min)  \\
%\hline
%  $l1$  & 260   \\
%  $l2$  & 300   \\
%  $l3$  & 340   \\
%  $h1$  & 270   \\
%  $h2$  & 250   \\
%  $h3$  & 380   \\
%  $r1$  & 370   \\
%  $r2$  & 390   \\
%\hline
%\end{tabular}
%\tablenotetext{a}{Footnote text here.}
%\end{table}

% See below for how to make landscape/sideways figures or tables.

\end{document}